\documentclass{mpe_report}

\usepackage{psfig,graphicx,epsfig}
\usepackage{color}
\usepackage{amsmath,amssymb,epic,eepic,array}

\begin{document}

\pagenumbering{arabic}
\setcounter{page}{153}

 \renewcommand{\FirstPageOfPaper }{153}\renewcommand{\LastPageOfPaper }{156}
\title{Evolution of the Inclination Angle of Radio Pulsars is Observable Effect}
\author{Ya. N. Istomin\inst{1} \and T.V. Shabanova\inst{2}}  
\institute{Theoretical Department, P. N. Lebedev Physical Institute, Leninsky Prospect 53, 
119991 Moscow, Russia \and
Astro Space Center, P. N. Lebedev Physical Institute, 
119991 Moscow, Russia}

\maketitle

\begin{abstract}

It is shown that the slow glitches in the spin rate of the pulsar B1822-09
can be explained by the reconstruction of the neutron star shape, which is not
matched with the star rotation axis. Owing to the evolution of the inclination angle, i.e.
the angle between the rotation axis and the axis of the magnetic dipole, under the action
of the braking torque, there appears the disagreement between the rotation axis and the symmetry
axis. After the angle between the axis of symmetry and the axis of the rotation achieves 
the maximum value of $\alpha\simeq 2\cdot 10^{-4}$ the shape of
the neutron star becomes matching with the rotation axis. Such reconstruction is observed as
the slow glitch. 

\end{abstract}

\section{Introduction}

All radio pulsars are known to retard theirs rotation. Periods of the rotation $P$ are gradually
increasing with the rate of the order of  
$\dot{P} \approx 10^{-15} s/s$. As a result during the time  
${P}/{\dot{P}} \approx 10^{7} years$ the rotation period becomes twice larger. At the same time the
inclination angle $\chi$ -- the angle between the rotation axis and the axis of the magnetic dipole --
must changes also. The characteristic time of the evolution of the $\chi$ is the same as for the rotation
period $P$. The change of the inclination angle is of $\approx 1$ during the evolution time.
Change of the value of $\chi$ is due to the torque acting on the star is not parallel to the rotation axis.
This torque not only retards the rotation but also turns the rotation axis in space.   
Increasing or decreasing of the $\chi$ depends on the mechanism of a neutron star energy losses.
The angle $\chi$ approaches to the value when the energy losses are minimal.
For the magnetodipole losses, if a pulsar radiates the electromagnetic magnetodipole waves, the angle
$\chi$ goes to the zero value when the radiation is absent. Opposite, for the current losses, if a
pulsar spends its energy for the creation and acceleration of plasma in a magnetosphere, and a star's spin 
down is due to the electric currents flowing in a magnetosphere and closing on a star's surface, the
angle $\chi$ approaches to the value of ${90}^{\circ}$. In this case the electric current is minimal.

However, it is clear that to notice during the observation time $\approx$ 10--20 years
the change of the inclination on the value of $10^{-6} rad\approx {0.}^{\prime\prime}2$ 
is impossible. Thus, there exist only indirect conclusions about inclination angle
evolution. They are based on the statistic of the distribution of pulsars over the value of
$\chi$. But because we do not know this distribution at the birth time, and due to the selection
effects we can't now do the definite conclusion about inclination angle evolution.    

Our work demonstrates the possibility to watch the changes of $\chi$ on the value of $2^{\prime}$ 
during the time $\approx 10 years$ though slow glitches observed from the pulsar B1822-09.

\section{Observed properties of slow glitches from the pulsar B1822$-$09}

The pulsar B1822$-$09 has the period $P=0.769 s$  ($\nu=1/P=1.3 s^{-1}$),
the period derivative $\dot{P}=52.36\times10^{-15}$. It is the young pulsar with the age of
$P/{2\dot{P}} \approx 2.3\times10^{5} years$. The pulsar observations during the time of
21 years at Puschino Radio Observatory show two class of the period variations. They are
pulsar noise and glitches. The pulsar B1822-09 reveals the new kind of glitches unknown before --
slow glitches (Shabanova 2005).

\begin{figure}
\centerline{\psfig{file=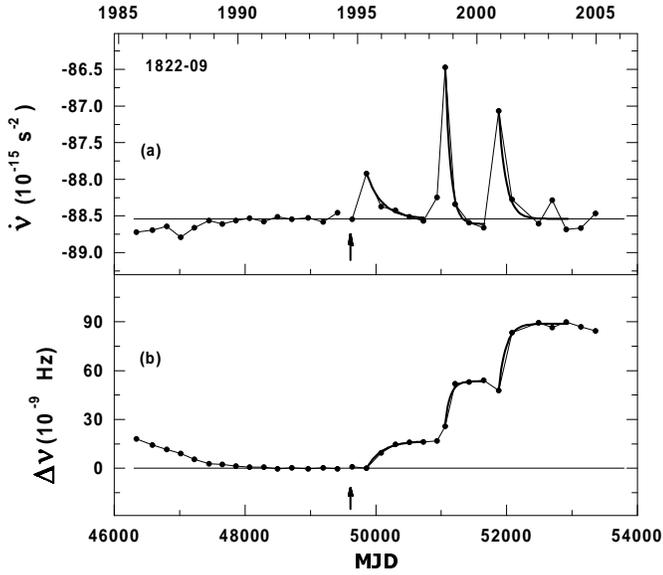,width=8.8cm,clip=} }
\caption{The time dependance (a) of the first derivative of the rotation frequency
${\dot\nu}$ and (b) frequency resiuals $\Delta{\nu}$ for the pulsar
B1822$-$09. Three slow glitches are observed from 1995 to 2005.
\label{image}}
\end{figure}

Figure 1 shows the frequency first derivative $\dot{\nu}$ and frequency residuals
$\Delta{\nu}$ for PSR B1822-09 from 1985 to 2006. The local fits were performed over intervals of 
200 days. The figure 1(b) presents the frequency residuals with respect to the simple spin
down model based on the initial parameters of $\nu$ and $\dot{\nu}$, determined from 1991 to 1994.
The narrow indicates the time at which the usual glitch of the magnitude  
$\Delta{\nu}/{\nu}=8\times10^{-10}$ was observed. This moment divides the observations on two parts.
After that time we observe three slow glitches.

During the interval from 1985 to 1994 we observed the distinct linear decreasing of the first derivative
of the period. The value of the second derivative is
$\ddot{\nu}=9\times10^{-25} s^{-3}$. 
The data from 1995 to 2006 show three jumps of values of $\dot{\nu}$ and $\Delta{\nu}$, 
which can be called as slow. The characteristic property of slow glitches is the gradual increasing
of the rotation frequency during 200-300 days without subsequent relaxation.
The increasing of the relative rotation frequency $\Delta{\nu}/{\nu}$ due to the slow glitch is of
$10^{-8}$. The total increasing for three glitches is
$\Delta{\nu}/{\nu} \approx 7\cdot 10^{-8}$.

We see that three slow glitches observed from PSR B1822-09 shows the quite different properties than
well known glitches from another pulsars, for example, from Crab and Vela pulsars.
They are connected with processes occurred in a core and a crust of a neutron star. But the reason of
the slow glitch also can't be connected with a magnetospheric effects.  
If a pulsar magnetosphere is responsible for a slow glitch then it must change its moment of inertia
$J$ on the value of $\Delta{J}/{J}=- \Delta{\nu}/{\nu} \approx -10^{-8}$. The pulsar magnetosphere
does not possess the such moment of inertia. The estimation of the maximum value of $J_m$ gives
$J_{m} < \int (B^{2}/8{\pi}c^{2})r^{2} d{\vec{r}}$.
Here $B$ is the magnitude of the magnetic field in a magnetosphere, $c$ is the speed of light,
$r$ is the distance from the rotation axis. The integration is over a pulsar's magnetosphere from
a star's surface to the light surface $c/{\Omega}=2{\pi}c/{\nu}$. As a result we obtain  
$J_{m}<B_{0}^{2}R^{5}/2c^{2} \approx 5\times10^{32}
{(B_{0}/10^{12}G)}^{2}{(R/10^{6}} cm)^{5} g cm^{2}$.
$R$ is the star's radius.
Thus, $J_{m}/J < 5\cdot 10^{-13}$, where $J \approx 10^{45} g cm^2$ is the standard value of the
moment inertia of a neutron star. So, the changes of the rotation frequency on the value of
$\Delta{\nu}/{\nu}\approx 10^{-8}$ as observed can be only by changes in a star's body.

\section{Connection between the inclination angle and slow glitches}

\subsection{Reconstruction of a star's shape}
The neutron star shape has no the spherical symmetry. Due to the fast rotation its extension across
the rotation axis is larger than that along. As a result the tensor of inertia of a star 
$J_{\alpha\beta}$ has no equaled diagonal components. The value of 
$J_{zz}$ (the axis $z$ is directed along the rotation axis) is little larger than the components
which are perpendicular to the rotation axis $J_{\perp\perp}$: $J_{zz}=J$, 
$J_{\perp\perp}=J-{\delta}J$, ${\delta}J {\ll}J$.
The ratio ${\delta}J/J$ is of the order of the ratio of the centrifugal force to the gravitation force 
$$\frac{{\delta}J}{J} \approx \frac{\Omega^{2}}{G{\rho}_{n}}.$$
Here $\Omega$ is the frequency of a star's rotation, $G$ is the gravitation constant,
$\rho_{n}$ is a star's density. Substituting the standard value of a neutron star density
$\bar{\rho}=5\times10^{14} g cm^{-3}$, we obtain  
$$\frac{\delta{J}}{J} \approx 1.2\times10^{-6}{P^{-2}}
\left(\frac{\bar{\rho}}{\rho_{n}}\right).$$

If the inclination angle changes then the rotation axis deflects from the symmetry axis
$z$ (the magnetic field is frozen to the star's body). The axis of rotation begins to precess around
the symmetry axis. That is described by relations
$\Omega_{x}=\Omega\sin{\alpha}\cos{\Phi}$;
$\Omega_{y}=\Omega\sin{\alpha}\sin{\Phi}$;
$\Omega_{z}=\Omega\cos{\alpha}=const(t)$;
$d{\Phi}/dt=\Omega{\cos{\alpha}}({\delta}J/J)$, 
from which the period of precession $P_{pr}$ follows
$P_{pr}=P(J/{\delta}J)/{\cos{\alpha}} \approx 0.83\cdot 10^{6}{P^{3}}({\rho_{n}}/{\bar{\rho}})
({\cos{\alpha}})^{-1} s$. For the pulsar B1822-09, which has the period $P=0.769 s$,
the precession period is equal to ($\alpha{\ll}1$):
$P_{pr} \approx 3.8\times10^{5} s$ = 4.4 days. This time is significantly less than the pulsar's age
$P/{\dot{P}}=1.5\times10^{13} s = 5\times10^{5} years$ and the observation time. Averaging over
the precession period $P_{pr}$, we get $<{\vec{\Omega}}>=\Omega_{z}{\vec{e}}_{z}$. The observed period is
equal to $P=2{\pi}/\Omega_{z}$.

The torque acting on star results the spin down of a star ($\dot{P}>0$) and also changing of the 
inclination angle $\chi$. As a result the angle $\alpha$ increases. The star's shape, initially
matched with the rotation axis, becomes asymmetric with respect to the instant axis of rotation.
The star, rotating around the axis not coinciding with the axis of symmetry, undergoes the stress. 
This stress intends to reconstruct the star's shape in order to match it with the rotation axis.
At $\alpha\ne 0$ the part of the centrifugal force is not compensating by the gravitation force.
So, there exists the force directing tangent to the surface of constant density (see figure 2).

\begin{figure}
\centerline{\psfig{file=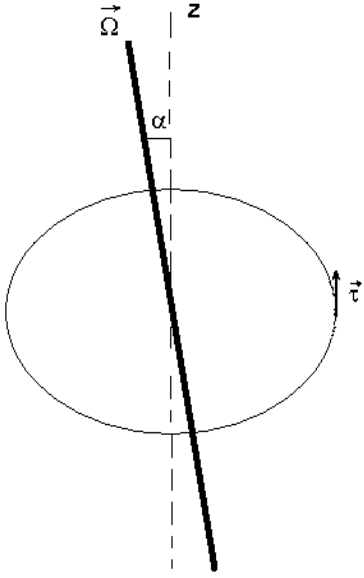,width=5.0cm,height=8.0cm,clip=} }
\caption{The star rotates around the axis not coinciding with the symmetry axis. The angle 
between axises is $\alpha$. There arises the tangent stress ${\bf \tau}$ intending to reconstruct the star's
shape.
\label{image}}
\end{figure}

The force density $f$, acting on the unit volume, is $f=\rho_{n}\Omega^{2}r{\alpha}$, where 
$r$ is the distance from the axis of rotation. The arising tangent stress equals
$\tau=F/s=\alpha{\Omega}^{2}
{\int\limits_0^h \rho_{n}(h^{\prime})(R-h^{\prime})dh^{\prime}}$
(here $R$ is the star's radius). The stress $\tau$ will intend to deflect the star matter column on
the angle $\phi$, 
$${\phi}(h) \approx \alpha{\Omega}^{2}
{\int\limits_0^h \frac{\rho_{n}(h^{\prime})}{E(h^{\prime})}
(R-h^{\prime})dh^{\prime}}.$$

The quantity $E$ is the elastic module, it depends on the matter density $\rho_{n}$, 
$E=E_{s}(\rho_{n}/{\rho_{s}})^{4/3}$.
Quantities $\rho_{s}$ and $E_{s}$ are values of the density and the elastic module
on the star's surface. The neutron star crust near the surface consists of the crystallic iron.
Its density on the star surface is of
$\rho_{s} \approx 10^{5} g/cm^{3}$. The density increases with the depth $h$, 
$\rho_{n}=\rho_{s}[1+(h/H)^{3}]$. The quantity $H$ is the characteristic scale of a crust density growth, 
$H\approx10 m$ (Blandford, Applegate, Hernquist, 1983). The value of the elastic module $E_s$ 
can be estimated from the magnitude of $E$ for
a steel on the Earth,
$E_{0} \approx 2\times10^{12} din/cm^{2}$. Then
$E_{s}=E_{0}(\rho_{s}/\rho_{0})^{4/3} \approx 3\cdot 10^{17} din/cm^{2}$
(at $\rho_{0} \approx10 g/cm^{3}$). 
Resulting we obtain the following expression for the angle of deflection of star matter from the vertical
$$\phi(h) \approx \alpha\frac{\Omega^{2}\rho_{s}}{E_{s}}
\int\limits_0^h \frac{(R-h^{\prime})dh^{\prime}}
{[1+(h^{\prime}/H)^{3}]^{1/3}}.$$

At $h{\ge}H$ the angle of deformation approaches the maximum value of
$\phi\to \alpha(\Omega^{2}{\rho_{s}}RH/{E_{s}})$,
which is much less than the angle $\alpha$,
$\phi \approx 3\times10^{-2}\alpha$.
Thus, we see that the elastic deformation is not enough to reconstruct the shape of a star to the 
equilibrium form.

While the angle $\alpha$ grows the tangent stress $\tau$ increases and can become greater than the plastic
limit for the iron crust $\sigma(\rho_{n})$. The elastic limit $\sigma$ is proportional to the value of
$E$ and is of the small part of it,
$\sigma(\rho_{n})=\sigma_{s}(\rho_{n}/\rho_{s})^{4/3}$
($\sigma \approx 10^{-3}E$ on the Earth at the room temperature).
The value $\sigma_{s}$ is the elastic limit for the neutron star crust at its surface.
The condition $\tau=\sigma$ defines the limiting value of the angle 
$\alpha_{l}$ at which the crust matter will move under the action of stress. After that the star shape
becomes symmetric with respect the rotation axis.
$$\alpha_{l}=\frac{\sigma_{s}}{\Omega^{2}\rho_{s}RH}
\frac{(1+x^{3})^{4/3}}{\int\limits_0^x (1-{\beta}x^{\prime})
(1+{x^{\prime}}^{3})dx^{\prime}},$$
where $x=h/H,\, \beta=H/R \approx10^{-3}{\ll} 1.$
The right hand side of this expression has the minimum at
$x=2^{-1/3}$. This minimum does define the limiting value of the angle 
$\alpha_{cr}$. At this value of $\alpha$ the star will reconstruct its shape, and the matter motion
will be mainly at the depth 
$h \approx 0.8H$,
\begin{equation}
\alpha_{cr} = 1.9\frac{\sigma_{s}}{\Omega^{2}{\rho_{s}}RH}.
\end{equation}

The value of the plastic limit $\sigma$ depends strongly on the matter temperature. More exactly,
on the ratio of the binding energy of atoms to the temperature. This ratio is proportional
to the quantity ${\rho_{n}}^{1/3}/T$. The matter density on the star's surface 
$\rho_{s} \approx10^{5} g/cm^{3}$ is much greater than that for metals on the 
Earth $\rho_{0} \approx10 g/cm^{3}$. The star surface temperature is also much higher, $T \approx 10^{5}\,K$.
So, the parameter $\rho_{s}^{1/3}/T$ is of the order less than that for the Earth conditions,
$\rho_{0}^{1/3}/T_{0} \, (T_{0} \approx300\,K)$, namely
$\rho_{s}^{1/3}/T \approx 6.5\times10^{-2}
(\rho_{0}^{1/3}/T_{0})(T/10^{5}\,K)^{-1}$.
Such value of the binding parameter corresponds to the strongly heating body, for which
the plastic limit $\sigma$ is at least of the order less than that for the room temperature.
Because of that we adopt the following value for $\sigma_{s}$, 
$\sigma_{s} \approx 10^{12} din/cm^{2}$. The obtained estimation of the critical value of the
angle between the axis of rotation and the symmetry axis for the pulsar B1822-09 is
$\alpha_{cr} \approx 2\times10^{-4}$.
The angle $\alpha$ can not exceed this limit.
During the evolution of the inclination angle $\chi$ the angle $\alpha$ increases from zero to
$\alpha_{cr}$. Then, the crust reconstructs, and the angle $\alpha$ returns to zero value.
This reconstruction is observed as a slow glitch.

\subsection{Change of the star rotation frequency after the glitch}

Let us discuss the behavior the star rotation frequency after the glitch.
The value of $\Omega$ after the reconstruction 
$\vec{\Omega}^{\prime}$ can be determined form the condition of angular momentum conservation
$J_{\alpha\beta}{\Omega}_{\beta}=const$.
After the glitch the angular velocity is directed along the symmetry axis, i.e.
$J_{\alpha\beta}{\Omega_{\beta}}^{\prime}=J{\Omega_{\alpha}}^{\prime}$.
Thus,
${\Omega_{x}}^{\prime}=\Omega(1-{\delta}J/J)\sin{\alpha}\cos{\Phi};$
${\Omega_{y}}^{\prime}=\Omega(1-{\delta}J/J)\sin{\alpha}\sin{\Phi};$
${\Omega_{z}}^{\prime}=\Omega\cos{\alpha}.$
Here the quantity $\Omega$ is the star rotation frequency before the glitch.
Let remind that the mean rotation frequency before the glitch is
$<\Omega>=2{\pi}/P=\Omega\cos{\alpha}$.
Calculating of ${\Omega}^{\prime}$, we find the change of the star rotation
$\Delta{\Omega}=\Omega^{\prime}-<\Omega>=\Omega(\alpha^{2}/2)
(1-2{\delta}J/J) \approx \Omega(\alpha^{2}/2)$.
So,
$$\frac{\Delta{\Omega}}{\Omega}=\frac{\alpha_{cr}^{2}}{2}>0.$$
Indeed, the observed jumps 
$\Delta{\Omega}$ for PSR B1822$-$09 are positive and are of the order of
$\Delta{\Omega}/{\Omega} \approx 10^{-8}$. These values are consistent with the estimation
$\alpha_{cr} \approx 10^{-4}$, obtained above.

\subsection{The rate of the inclination angle change}
To achieve the value
$\alpha \approx10^{-4}=21^{\prime\prime}$,
and also the change of ${\chi}$ on the same value it is necessary large enough time
$t\approx(P/\dot{P}){\alpha}$.  For a usual pulsar this time is of
$3\times10^{3} years$.
PSR B1822$-$09 possesses the much higher value of the spin down
$\dot{P}=52.36\times10^{-15}$, and the corresponding time diminishes to
$t\approx 50 years$. That is also large for the observations. But the pulsar
B1822$-$09 is on their stage of evolution that this time becomes shorter and is of several years.
The matter is while approaching the value of inclination angle $\chi$ to the limiting value 
(0 or ${\pi}/2$ depending on the mechanism of losses), when the
energy losses become minimal,
the change of $\chi$ becomes larger than that of $\Omega$. For the magnetodipole losses the invariant is
the quantity
$\Omega\cos{\chi}=const$. It gives the rate of $\chi$, $\dot{\chi}=-(\dot{P}/P)\cot{\chi}$.
At the angle $\chi$ closed to zero, $\cot{\chi}{\gg}1$ and ${\dot \chi}\gg{\dot\Omega}$.
The time of changing of $\chi$ on the value of $\Delta{\chi}$ equals  
$t=(P/\dot{P})\Delta{\chi}\tan{\chi}{\ll}(P/\dot{P})\Delta{\chi}$.
The same relation takes place also for the current losses. The invariant value for this case is
$\Omega\sin{\chi}=const \, (\chi\ne{\pi}/2)$.  The angle $\chi$ approaches ${\pi}/2$. 
The time is
$t=(P/\dot{P})\Delta{\chi}\cot{\chi}$, and it is also smaller than
$(P/{\dot{P}})\Delta{\chi}$ at $\chi \approx {\pi}/2$.

Existence of the interpulse in radio emission of PSR B1822-09 tell us that the value of 
$\chi$ is close to $0$ or to ${\pi}/2$. 
Rankin (1990) determined this angle,
$\chi \approx 86^{\circ}$. The conclusion that the angle $\chi$ is close to the value of ${\pi}/2$
is confirmed by the fact that the shift between the main pulse and the interpulse is equal to 
$180^{\circ}$ and does not depends on the radio frequency (Gil et al., 1994). 

One more argument in favor of orthogonal rotator
($\chi \approx {\pi}/2$) is the measuring braking index
$n=\Omega{\ddot{\Omega}}/{\dot{\Omega}}^{2}$ has the large value of
$n \approx 145$. Such high value of the braking index, $n{\gg}1$, means that the rate of change of the 
inclination angle $\chi$ is much larger than that of the rotation frequency $\Omega$. Taking into 
account only the rotation frequency evolution it is impossible to get such high value of $n$, $n \approx 3$. 
In the relation
$\dot{\Omega}=f(\Omega{,}\chi)$ at $n{\gg}1$ the dependence on $\chi$ becomes determinant
$\ddot{\Omega} \approx \dot{\chi}{\partial}{f}/{\partial}{\chi}$.
By this the braking index is
$n \approx \dot{\chi}(\Omega/\dot{\Omega})(\partial\ln{f}/\partial{\chi})$.
Thus, the braking index is defined by
${\dot\chi} \approx n({\dot\Omega}/{\Omega})
(\partial\ln{f}/\partial{\chi})^{-1}{\gg}{\dot{\Omega}}/\Omega$
at $n{\gg}1$.

For the current losses which give the evolution of the inclination angle to ${\pi}/2$,
$f(\chi)\propto \, {\cos}^{2}{\chi}+{\delta^{2}}/4$ (Beskin, Gurevich, Istomin, 1993).
Here the quantity $\delta$ is the angular size of the polar cap,
$\delta=(1.95R{\Omega}/c)^{1/2}=2.3\cdot 10^{-2}(R/10 km)^{1/2}
\approx {1.3}^{\circ}$. The value of ${\dot \chi}$ has the sharp maximum near
$\cos{\chi} \approx {\delta}/2$, at which
$\dot{\chi}=-({\dot \Omega}/\Omega)\cos{\chi}/
(\cos^{2}{\chi}+{\delta}^{2}/4)$. 
The characteristic value of 
${\dot\chi}$ in the interval of angles ${\pi}/2-\chi \approx \delta$ is
$\dot{\chi}=-\delta^{-1}({\dot{\Omega}}/{\Omega})=\delta^{-1}{\dot{P}}/P
\approx 43{\dot P}/P$. At such high rate the change of $\chi$ on the value of
$\Delta{\chi} \approx 10^{-4}$ is during the time scale of several years.
We suggest that the pulsar B1822-09 demonstrates such kind of evolution.

\subsection{Comparison with observations}
We consider that observed slow glitches of B1822-09 are reconstructions of the neutron star shape.
We observed after 1995 three slow glitches of the pulsar rotation,
$\Delta{\nu}_{1}/\nu=1.2\times10^{-8}$,
$\Delta{\nu}_{2}/\nu=2.85\times10^{-8}$,
$\Delta{\nu}_{3}/\nu=2.75\times10^{-8}$.
Knowing the relation $\alpha^ {2} =2{\Delta {\nu}}/\nu$, we obtain the following estimations for the angles
$\alpha_{1}=1.5\times10^{-4}\, (\approx30^{\prime\prime})$,
$\alpha_{2}=2.4\times10^{-4}\, (\approx50^{\prime\prime})$,
$\alpha_{3}=2.3\times10^{-4}\, (\approx48^{\prime\prime})$.
We see that observed values of
$\alpha$ are close to the critical one
$\alpha_{cr}=2\times10^{-4}$, derived above from the condition of plastic limit.
To achieve these angles we need the time of
$t=\alpha/{\dot\chi}$.

For the orthogonal rotator ($\chi \approx {\pi}/2$) the quantity ${\dot\chi}$ is
${\dot\chi}=\delta^{-1}{\dot P}/P \approx 2.9\times10^{-12} s^{-1}$, and the time to achieve
$\alpha=2\times10^{-4}$ is
$t=7\times10^{7} s=800 days$. This time is in well agreement with observed intervals 
$t_{2}\approx t_{3}\approx 700 days$.

Thus, we come to the conclusion that slow glitches observed from PSR
B1822$-$09 is explained by the model of the neutron star reconstruction, when the asymmetry
of the star's shape is restored by the plastic deformation in the crust.
We also showed the pulsar B1822-09 is the orthogonal rotator, and $\chi\approx 90^{\circ}$. 
This is the confirmation of the current losses conception.

\section{Conclusion}

We showed that the probable reason of slow glitches of the pulsar 
B1822$-$09 is the fast change of its inclination angle $\chi$. 
During slow glitches the inclination angle changes on the value of
$2^{\prime}$ for 10 years.
Significant increasing of the rate
${\dot\chi}$ takes place when the angle
$\chi$ approaches $90^{\circ}$, and the condition
${\pi}/2-\chi {\approx}\delta$ is fulfilled ($\delta$ is
the angular size of the polar cap $\delta\approx (2R{\Omega}/c)^{1/2}$).
In this region of angles 
$\dot{\chi}={\delta}^{-1}{\dot{P}}/P{\gg}{\dot{P}}/P$.

For the fast evolving pulsars like Crab, Vela or B1509$-$58,
for which the ratio ${\dot{P}}_{-15}/P$ is large quantity of $10^{3}-10^{4}$, 
the time of accumulation of angle $\alpha_{cr}$ is of the order of one year.
They have to demonstrate slow glitches. But they have came to the state when
$\chi= 90^{\circ}$. For another group of pulsars with the ratio ${\dot{P}}_{-15}/P {\approx}10^{2}$ 
the accumulation time is of 30 years. And only for the pulsar B1822-09, which has the angle $\chi$ close
to $90^{\circ}$, but not equal that value, the accumulation time is of several years.

          \clearpage

\end{document}